\documentclass{elsart3}

\usepackage{graphicx}
\usepackage{amssymb}

\begin{document}

\begin{frontmatter}

\title{Phase transitions in ferro-antiferromagnetic bilayers with a stepped
interface}

\author[ad1]{D.P. Landau\corauthref{cor1}},
\corauth[cor1]{Corresponding author. Tel +1 (706) 542 2908, Fax +1 (706) 542 2492}
\ead{dlandau@uga.edu}
\author[ad1,ad3]{Shan-Ho Tsai},
\author[ad2]{Thomas C. Schulthess}
\address[ad1]{Center for Simulational Physics, University of Georgia, Athens, GA 30602}
\address[ad2]{Center for Computational Sciences and Computer Science \&
Mathematics Division, Oak Ridge National Laboratory, Oak Ridge, TN 37831}
\address[ad3]{Enterprise IT Services, University of Georgia, Athens, GA 30602}

\begin{abstract}
We have studied magnetic ordering in ferro/antiferromagnetic (F/AF) 
bilayers using Monte Carlo simulations of classical Heisenberg spins. 
For both flat and stepped interfaces we observed order in the AF 
above the N\'eel temperature, with the AF spins aligning collinearly 
with the F moments. In the case of the stepped interface there is a 
transition from collinear to perpendicular alignment of the F and AF 
spins at a lower temperature.

\end{abstract}

\end{frontmatter}

Magnetic properties of ferro/antiferromagnetic (F/AF) 
bilayers can be very different from those of free F and AF films. 
These coupled systems exhibit a shift in the 
hysteresis loop (exchange bias) and larger coercivity than the free F film
\cite{reviews}.
Other effects of the F/AF coupling include the order in the AF 
observed above the N\'eel temperature\cite{vanderzaag00}
and the perpendicular orientation of the F with the AF 
moments\cite{perpthy,perp}. Understanding these effects and the nature 
of the F/AF interfaces remains a challenge.

In this paper we use Monte Carlo simulations to study the effect of
the interfacial exchange and roughness on the magnetic ordering in 
F/AF bilayers. The work is motivated by recent experiments for 
Fe$_3$O$_4$/CoO multilayers\cite{vanderzaag00}.
We consider a ferromagnetic (F) film coupled to an 
antiferromagnetic (AF) film where the lattice is coherent across the 
F/AF interface. Each film is a bcc lattice, with linear sizes
$L_x=L_y=L\le 96$ and $12$ staggered (because of the bcc structure) 
layers of classical spins $|{\bf S_r}|=1$, which interact
via the Hamiltonian
$${\mathcal H}=-J_F\!\!\!\!\!\sum_{\langle{\bf r},{\bf r'}\rangle\in {\rm F}}\!\!\!\!\!{\bf S_r}\cdot {\bf S_{r'}}
-K_F\!\!\sum_{{\bf r}\in {\rm F}}\!(S_{\bf r}^z)^2 - $$
$$-J_A\!\!\!\!\!\!\sum_{\langle{\bf r},{\bf r'}\rangle\in {\rm AF}}\!\!\!\!\!{\bf S_r}\cdot {\bf S_{r'}}
-K_A\!\!\sum_{{\bf r}\in {\rm AF}}\!\!(S_{\bf r}^y)^2 -J_I\!\!\!\!\!\!\sum_{\langle{\bf r},{\bf r'}\rangle\in {\rm F/AF}}\!\!\!\!\!{\bf S_r}\cdot {\bf S_{r'}}$$
where $\langle{\bf r},{\bf r}'\rangle$ denotes nearest-neighbor pairs of spins
coupled with exchange interactions $J_F=5J>0$ in the F film, $J_A=-J<0$ in the
AF film, and $J_I=-J$ at the F/AF interface.
Spins in the AF film have a uniaxial single-site anisotropy $K_A=J$, whose
easy axis is along the $y$ axis. The demagnetizing field on the F film is
modeled with a hard-axis ($K_F=-0.5J$) along the $z$ direction,
which is perpendicular to the F/AF interfacial plane. No external magnetic
field is applied.
We use periodic boundary conditions along the $x$ and $y$ directions and
free boundary conditions along the $z$ direction.
We model flat interfaces that are fully uncompensated as well as uniformily
stepped ones that are compensated on average. The stepped interface has
$6$, $L$, and one spin per terrace in the $x$, $y$, and $z$ directions, 
respectively. 

The F and AF order parameters are the uniform ($m$) and staggered ($m_s$) 
magnetization per spin, respectively\cite{mmm2002}. $m_x$ and $m_y$
denote the two components of $m$ in the interfacial plane. 
We perform Monte Carlo simulations with Metropolis algorithm %\cite{MCbook} 
at fixed temperature $T$. Typically we discard $3 \times 10^5$
Monte Carlo Steps/site (MCS) for thermalization and then use about
$2\times 10^5$ MCS for averages. Whenever not shown, error bars in the 
figures are smaller than the symbol sizes.

\begin{figure}[ht]
\centering
\leavevmode
\includegraphics[clip,angle=0,width=0.42\textwidth]{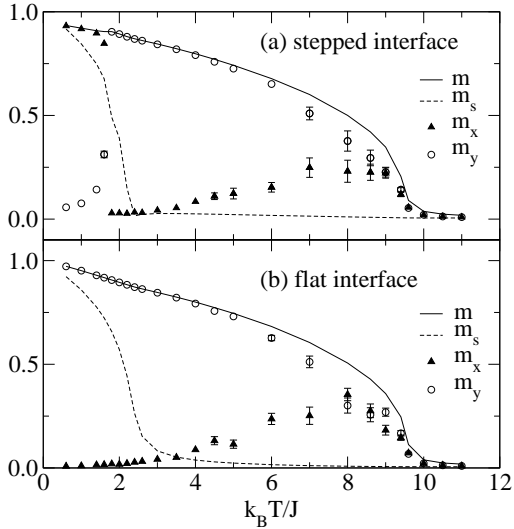}
\caption{\label{figcomp}Temperature dependence of $m$, $m_s$, $m_x$, and $m_y$ (see text) for $L=60$, with (a) stepped and (b) flat interface.}
\end{figure}

The $T$-dependence of $m$ shown in Figs.\ref{figcomp}(a) and 
\ref{figcomp}(b) plus finite-size analysis (not shown) indicate an F 
ordering transition at $T_c\approx 9.3J/k_B$.
A free 12-atomic-layer AF film undergoes a phase transition at the N\'eel 
temperature\cite{mmm2001} $T_N\approx 2.2J/k_B$. 
For both flat and stepped interfaces, the spins on the F film orient
predominantly in a direction collinear with the easy axis of the
AF at high $T$ (see Fig.\ref{figcomp}), even above $T_N$.
This is an indication
that there is still order in the AF above $T_N$ due to the
coupling to the ferromagnet. As $T$ is lowered in the
case of the stepped interface the F spins switch to orient in a
direction that is perpendicular to the AF spins. Our results suggest 
that the onset of this perpendicular orientation is very sharp and 
it occurs at a temperature below $T_N$. The $z$-components of $m$ and $m_s$
are very small for all $T$.
In the absence of the AF film, spins on the F film
have global rotation symmetry in the $x$-$y$ plane, which is their
easy plane. The preferential orientations
of the F spins observed either below or above $T_N$ result
from the exchange coupling to the AF film.

\begin{figure}[ht]
\centering
\leavevmode
\includegraphics[clip,angle=0,width=0.42\textwidth]{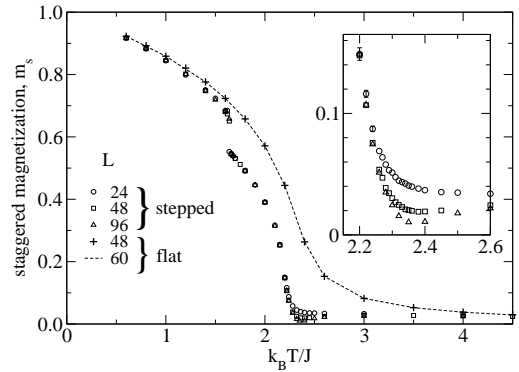}
\caption{\label{figstg}Staggered magnetization as a function of temperature for stepped and flat interfaces.}
\end{figure}

Fig.\ref{figstg} shows $m_s$ versus $T$ for different $L$ for stepped
and flat interfaces. In the former case the decay to zero of $m_s$ at 
$T\approx 2.2J/k_B$ becomes sharper for larger $L$ (see the inset), 
suggesting an AF phase transition at $T_N$.
In contrast, with a flat (uncompensated) interface
there is no finite-size dependency of $m_s$ near $T_N$, suggesting the
absence of an AF phase transition there.

Research partially supported by NSF grant DMR-0094422
and DOE-OS/ASCR through the Laboratory Technology Research Program 
under contract DE-AC05-00OR22725 with UT-Battelle, LLC.
Simulations performed on the IBM SP at NERSC and ORNL.

\end{document}